\documentclass[preprint,pra,aps,nofootinbib,longbibliography]{revtex4-1}
\usepackage{graphicx}

\begin{document}

\title{Fermion exchange in ring polymer self-consistent field theory}
\author{Malcolm A. Kealey}
\affiliation{Department of Physics \& Astronomy and Waterloo Institute for Nanotechnology, University of Waterloo, 200 University Avenue West, Waterloo, Ontario, Canada N2L 3G1}
\author{Philip A. LeMaitre}
\affiliation{University of Innsbruck, Institute for Theoretical Physics, Technikerstr. 21a, A-6020 Innsbruck, Austria}
\author{Russell B. Thompson}
\email{thompson@uwaterloo.ca}
\affiliation{Department of Physics \& Astronomy and Waterloo Institute for Nanotechnology, University of Waterloo, 200 University Avenue West, Waterloo, Ontario, Canada N2L 3G1}
\date{\today}

\begin{abstract}

A mapping is made between fermion exchange and excluded volume in the quantum-classical isomorphism using polymer self-consistent field theory. Apart from exchange, quantum particles are known to  be exactly representable in classical statistical mechanics as ring polymers, with contours that are parametrized by the inverse thermal energy, often called the imaginary time. Evidence in support of a previously used approximation for fermion exchange in ring polymer self-consistent field theory is given, specifically, that the use of all-contour interactions in the mean field picture instead of equal imaginary time interactions is justified based on the symmetry of ring polymers. It is also shown that the removal of forbidden thermal trajectories, both those that violate excluded volume directly and those that represent topologically inaccessible microstates, is equivalent to antisymmetric exchange. The electron density of the beryllium atom is calculated with ring polymer self-consistent field theory ignoring classical correlations, and very good agreement is found with Hartree-Fock theory which also neglects Coulomb correlations. The total binding energies agree to within less than $6\%$, which while still far from chemical accuracy, is remarkable given that the field theory equations are derived from first principles with zero free parameters. The discrepancy between self-consistent field theory and Hartree-Fock theory is attributed to classical Coulomb self-interactions which are included in Hartree-Fock theory but not in self-consistent field theory. A potential method to improve the agreement by more accurately representing electron-electron self-interactions in self-consistent field theory is discussed, as are the implications for quantum foundations of the quantum-classical mapping between fermion exchange and thermal trajectory excluded volume.
\end{abstract}

\maketitle

\section{Introduction}
Modern quantum simulation methods often use a dimensional trick to exactly convert quantum many-body calculations into classical ones \cite{Ceperley1995, Roy1999b, Zeng2014}. This method, introduced by Feynman in 1953 \cite{Feynman1953a, Feynman1953b, Feynman1953c} and since referred to as the quantum-classical isomorphism \cite{Chandler1981}, treats the inverse temperature as a fictional dimension and so allows the statistical mechanics of a quantum system to be viewed as the classical statistical mechanics of a ring polymer system, where the term ``ring polymer'' refers to a mathematical contour that starts and ends at the same point, with the inverse temperature parametrizing the curve \cite{Ceperley1995, Thompson2019, Thompson2020, Thompson2022, Thompson2023, Fredrickson2023}. The inverse temperature is sometimes referred to as the imaginary time due to it being a Wick rotation of the quantum matrix element \cite{Feynman1953b}. This quantum-classical mapping is exact except for the lack of quantum exchange in the ring polymer partition function. The mathematics of boson exchange can be viewed classically as the merging of polymer rings into larger rings, including separation back into smaller rings \cite{Thompson2023, Fredrickson2023}, and so fits naturally into the quantum-classical isomorphism. For this reason, ring polymer simulations of quantum systems are primarily applied to bosons -- Feynman originated the method for explaining the superfluid transition in helium \cite{Feynman1953a, Feynman1953b, Feynman1953c}. To our knowledge, there is no similar quantum-classical mapping for fermions. Such a mapping would be extremely valuable, not only because it would allow for effective approximations to be made for many-body fermion computations within the quantum-classical isomorphism \cite{Thompson2020, LeMaitre2023a, LeMaitre2023b}, but also because of the implications for quantum foundations (see section \ref{sec-RandD} and references \cite{Thompson2022, Thompson2023}). In this paper, we present and justify a quantum-classical mapping for fermions within the quantum-classical isomorphism using polymer self-consistent field theory (SCFT). 

Self-consistent field theory (SCFT) is a mean field statistical mechanics formalism for classical coarse-grained polymers \cite{Matsen2006, Matsen2002, Fredrickson2006, Qiu2006, Schmid1998}, including ring polymers \cite{Kim2012, Qiu2011}. In recent years, it has been applied to study quantum systems using the quantum-classical isomorphism both in its mean field form \cite{Thompson2019, Thompson2020, Sillaste2022, Thompson2022,  Thompson2023, LeMaitre2023a, LeMaitre2023b} and including fluctuations through field-theoretic simulations (FTS) \cite{Delaney2020, Fredrickson2022, Fredrickson2023}. Only boson systems have been examined to date using FTS due to the difficulty of incorporating fermion exchange in the quantum-classical isomorphism. Feynman and Hibbs discussed the numerical problems caused by fermion exchange \cite{Feynman1965}, and as mentioned, there has been no classical interpretation along the lines of boson exchange. Despite this, mean field SCFT has been applied exclusively to fermion systems in the form of the electron densities of atoms \cite{Thompson2019, Thompson2020, LeMaitre2023a, LeMaitre2023b} and diatomic molecules \cite{Sillaste2022}. These applications have been based on a postulated equivalence between fermion exchange in three dimensions (real space) and excluded volume between ring polymer trajectories in four dimensions (real space plus the fictitious thermal dimension) \cite{Thompson2020, Thompson2022,  Thompson2023, LeMaitre2023a, LeMaitre2023b}.

An excluded volume assumption for the paths of quantum particles in thermal-space is not new. It is implicit in quantum simulations \cite{Ceperley1995} and explicit in Feynman's original work \cite{Feynman1953b}. Although these are boson systems, the assumption is independent of quantum symmetry -- any particle with mass should be expected to have excluded volume trajectories and so the assumption should be kept for electrons or other fermion systems. It does not necessarily follow that excluded volume in 4D is the same as fermion exchange in 3D, although from a quantum-classical perspective, the assumption is not unreasonable: a 3D fermion system is mathematically identical to a 4D classical ring polymer system except that the 3D system has fermion exchange but not excluded volume, whereas the 4D system has excluded volume but not exchange. Given that boson exchange has a classical interpretation in terms of merging and separating ring polymers, it is logical to expect a similar mapping for fermion exchange, and excluded volume seems an obvious choice. 

Evidence in support of the excluded volume postulate is also available: SCFT calculations which use the assumption give correct qualitative atomic shell structure \cite{Thompson2020, LeMaitre2023a, LeMaitre2023b} and molecular bonding \cite{Sillaste2022}. The SCFT predictions are also quantitatively quite good, although they do not come close to chemical accuracy \cite{LeMaitre2023a, LeMaitre2023b}. This is not surprising however, since implementing excluded volume accurately is notoriously difficult even in classical systems, so there will always be confounding factors in computationally testing the excluded volume hypothesis. Instead, one option is to turn to scaling theory, where for the case of a high density uniform electron gas, the excluded volume hypothesis correctly predicts the energy of the system to scale with the electron density following a 5/3 power, in agreement with Thomas-Fermi theory \cite{Thompson2020, LeMaitre2023a, Thomas1927, Fermi1927}. A further scaling theory correction term agrees with Dirac's exchange correction to the Thomas-Fermi energy, and scales with the density to the 4/3 power \cite{LeMaitre2023a, Dirac1930}. Another option is to make use of known analytic constraints on the electron density and fields \cite{Parr1989, Lieb2002} to test whether the excluded-volume hypothesis adheres to these constraints or not. In reference \cite{LeMaitre2023b}, a number of these constraints were tested numerically, and it was shown that the electron density does not violate any of them, even for the approximate implementation of the excluded-volume hypothesis. These results are either remarkable coincidences, or there is some reason to take the excluded volume postulate seriously. In particular, the agreement with Dirac exchange suggests a strong link between 4D excluded volume and 3D fermion exchange.

In this paper, we put the excluded volume-exchange hypothesis on a stronger footing, and give a mathematical justification for the equivalence. We give a numerical example of the beryllium atom using first principles SCFT with no free parameters which shows very good agreement with Hartree-Fock (HF) theory, and we explain remaining discrepancies between SCFT and HF in terms of residual classical approximations. We give some reasons why previous implementations using the excluded volume postulate failed to give perfect agreement with HF theory, and we show that some of those approximations are actually better than expected. 

\section{Summary of Theory}
We give only a brief synopsis of the relevant SCFT equations; further details and derivations can be found in references \cite{Thompson2019, Thompson2020, Sillaste2022, Thompson2022,  Thompson2023, LeMaitre2023a, LeMaitre2023b} for the quantum case, and in references \cite{Matsen2006, Matsen2002, Fredrickson2006, Qiu2006, Schmid1998} for SCFT applied to polymers. 

The SCFT equations for a system of $N$ quantum particles have the identical structure as those of a system of $N$ classical ring polymers. The quantum particle spatially inhomogeneous number density is 
\begin{equation}
n({\bf r},\beta) = \frac{N}{Q(\beta)} q({\bf r},{\bf r},\beta)  \label{dens1}  
\end{equation}
where $\beta = 1/k_BT$ with temperature $T$ and Boltzmann's constant $k_B$. $q({\bf r},{\bf r},\beta)$ is a real and non-negative propagator that is the diagonal of the solution to a modified diffusion equation
\begin{equation}
\frac{\partial q({\bf r}_0,{\bf r},s)}{\partial s} = \frac{\hbar^2}{2m} \nabla^2 q({\bf r}_0,{\bf r},s) - w({\bf r},\beta) q({\bf r}_0,{\bf r},s)  \label{diff1}
\end{equation}
subject to the influence of a field $w({\bf r},\beta)$, which contains all interactions between quantum particles, and the initial condition
\begin{equation}
q({\bf r}_0,{\bf r},0) = \delta({\bf r}-{\bf r}_0)   .  \label{init1}
\end{equation}
Equation (\ref{dens1}) is normalized by the single particle partition function
\begin{equation}
Q(\beta) = \int q({\bf r},{\bf r},\beta) d{\bf r}  .  \label{Q1}
\end{equation}

Previously \cite{Thompson2020, Thompson2023, LeMaitre2023a, LeMaitre2023b} , quantum exchange and the Pauli exclusion principle have been encoded in SCFT for fermions using a simple mean field excluded volume contact force based on repulsions between all polymer segments. That is, an Edwards excluded volume term \cite{Edwards1965, Doi2001} contributes a Pauli field $w_P({\bf r},\beta)$ that is linear in the density to the total field in the diffusion equation (\ref{diff1})
\begin{equation}
w_P({\bf r},\beta) = g^{-1}_0 \int n({\bf r},\beta) d{\bf r}  .  \label{wP1}
\end{equation}
The basic form (\ref{wP1}) should be modified to remove self-interactions following references \cite{LeMaitre2023a, LeMaitre2023b} but, in the interest of clarity, the simple form given here can be used for discussion purposes without loss of generality. The prefactor $g^{-1}_0$, which has units of an inverse density of states, sets the strength of the contact forces and is analogous to the Flory-Huggins parameter $\chi$ in polymer physics \cite{Matsen2002} and the excluded volume parameter $v$ in the work of Edwards \cite{Edwards1965, Doi2001}. Although equation (\ref{wP1}) gives semi-quantitatively correct results for the shell structure of atoms \cite{Thompson2020, LeMaitre2023a, LeMaitre2023b} when combined with an external potential (for atoms, the Coulomb attraction of a nucleus) and electron-electron interactions, it falls short of complete agreement with HF theory, which includes fermion exchange exactly. 

\section{Results \& Discussion} \label{sec-RandD}

Allowing contact force repulsion between all polymer segments is a coarse approximation -- it is known that, unlike for polymer systems, excluded volume in path integral simulations of quantum particles should be enforced only between contour values of equal imaginary time \cite{Fredrickson2023, Feynman1953b, Ceperley1995}. To accomplish this, one needs to keep track of the density at each imaginary time slice value $s$, rather than just the density at $s=\beta$. In other words, the field given by (\ref{wP1}) should be replaced by the set of fields
\begin{equation}
w_P({\bf r},s) = g^{-1}_0 \int n({\bf r},s) d{\bf r}  .  \label{wP2}
\end{equation}
Instead of a single field calculated only at $s=\beta$, a family of fields, one for each value of $0 \le s \le \beta$ is required. This in turn requires a family of densities $n({\bf r},s)$ for $0 \le s \le \beta$ to be computed. The formula for the density of segments at position ${\bf r}$ and imaginary time $s$ is \cite{Kim2012}
\begin{equation}
n({\bf r},s) = \frac{N}{Q(\beta)} \int q({\bf r},{\bf r}^\prime,s) q({\bf r},{\bf r}^\prime,\beta-s) d{\bf r}^\prime \label{dens2}  
\end{equation}
where 
\begin{equation}
Q(\beta) = \int \int q({\bf r},{\bf r}^\prime,s) q({\bf r},{\bf r}^\prime,\beta-s)d{\bf r} d{\bf r}^\prime .  \label{Q2}
\end{equation}
Equation (\ref{Q2}) is independent of $s$ in that any value of $s$ will give the same result for $Q(\beta)$. One can verify that choosing $s=\beta$ in (\ref{dens2}) and (\ref{Q2}) gives back (\ref{dens1}) and (\ref{Q1}) for the density at $s=\beta$. It is straightforward to calculate the densities at each value of $s$ using (\ref{dens2}) and (\ref{Q2}) and to find the Pauli potentials for each value of $s$ from equation (\ref{wP2}). 

We implemented this numerically for atomic systems by subdividing the $s$ contour into small intervals and using initial perturbations on the fields and densities to break the symmetry along the contour. Not surprisingly, the perturbations were found to die away giving results identical to those found with the simpler formulas (\ref{dens1}) and (\ref{wP1}) which ignore imaginary time-slices and calculate the Pauli potential and densities allowing all polymer segments to interact. This happens because there is nothing to maintain broken symmetry along the imaginary time direction. In an atom, for example, besides the Pauli potential, the Coulomb potentials of the ion and electron-electron interactions do not depend on $s$. Only the Pauli potential can depend on $s$ through its dependence on the density, and the density through the Pauli potential. Unlike systems such as block copolymers, where asymmetry in the molecular architecture allows for the breaking of spatial symmetry, there is no inherent symmetry breaking mechanism in the quantum particle trajectory -- for homogeneous ring polymers, including quantum particles, all segments of the ring are identical. Therefore, in the mean field approximation, all time slice densities give identical results. One is therefore free to choose any single value of $s$, such as $s=\beta$, meaning that formulas (\ref{dens1}) and (\ref{wP1}) are the correct mean field excluded volume expressions. 

\begin{figure}
\centering
\begin{tabular}{ccc}
\includegraphics[scale=0.15]{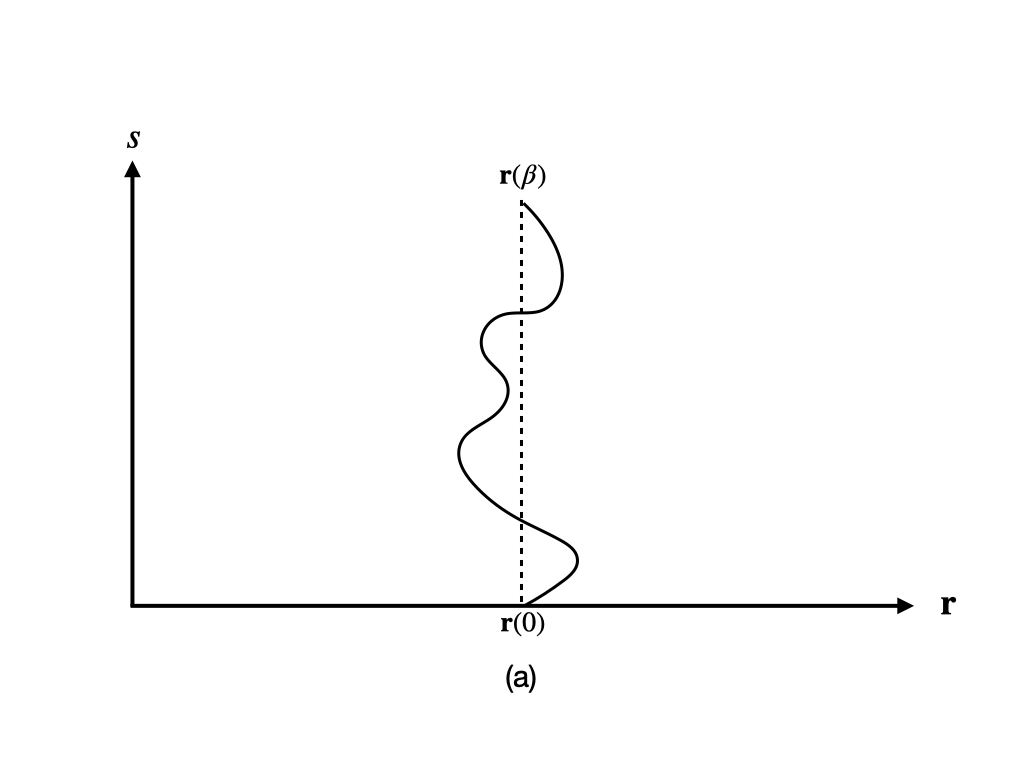} & \includegraphics[scale=0.15]{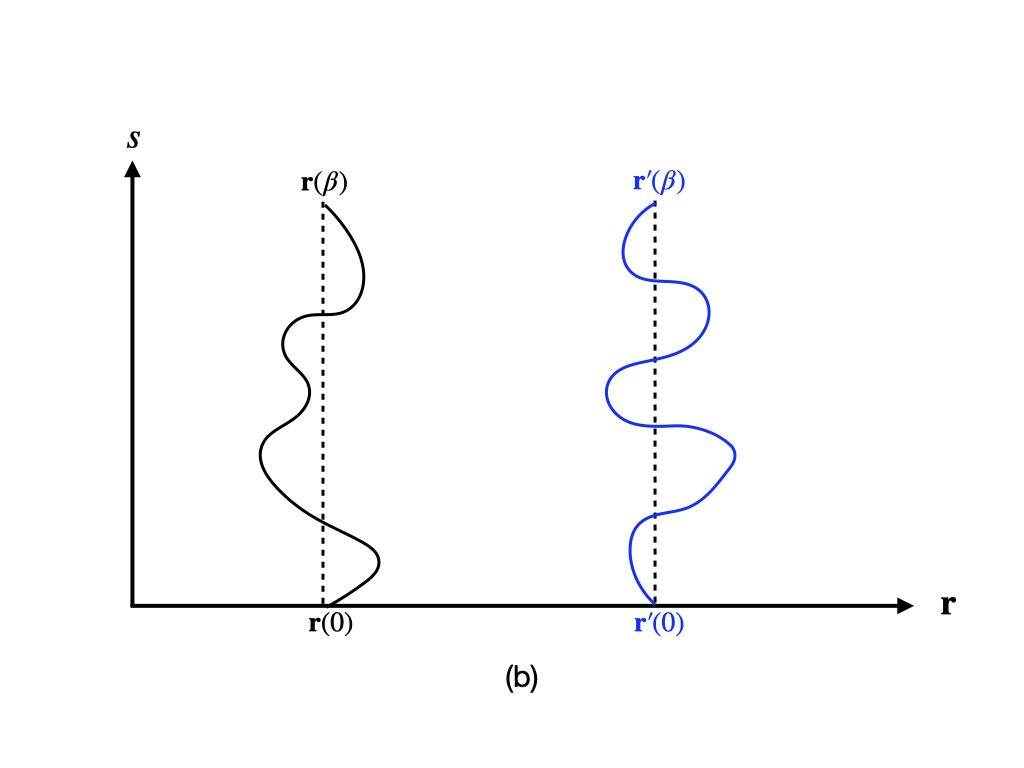} &
\includegraphics[scale=0.15]{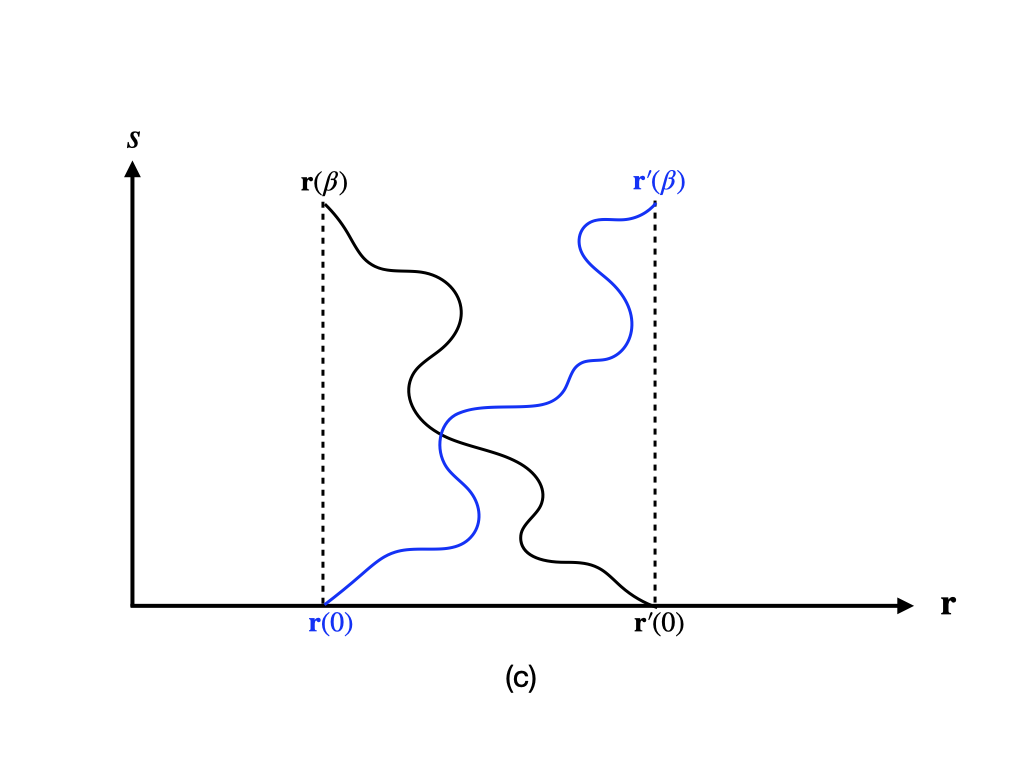} \\
\includegraphics[scale=0.15]{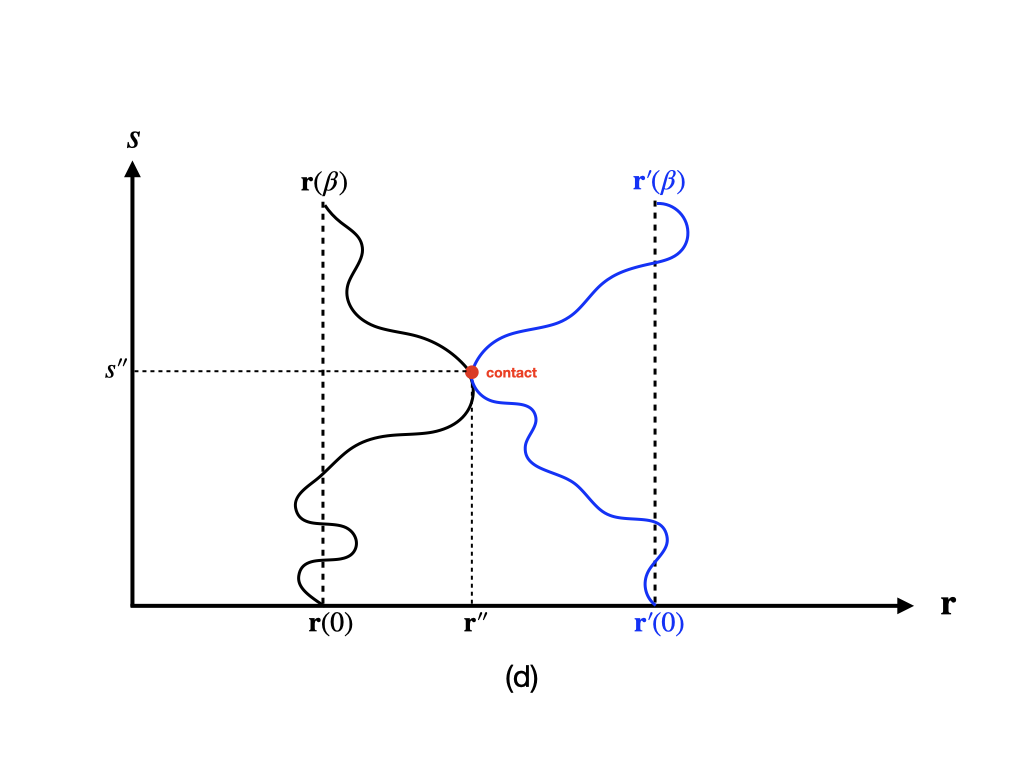} &
\includegraphics[scale=0.15]{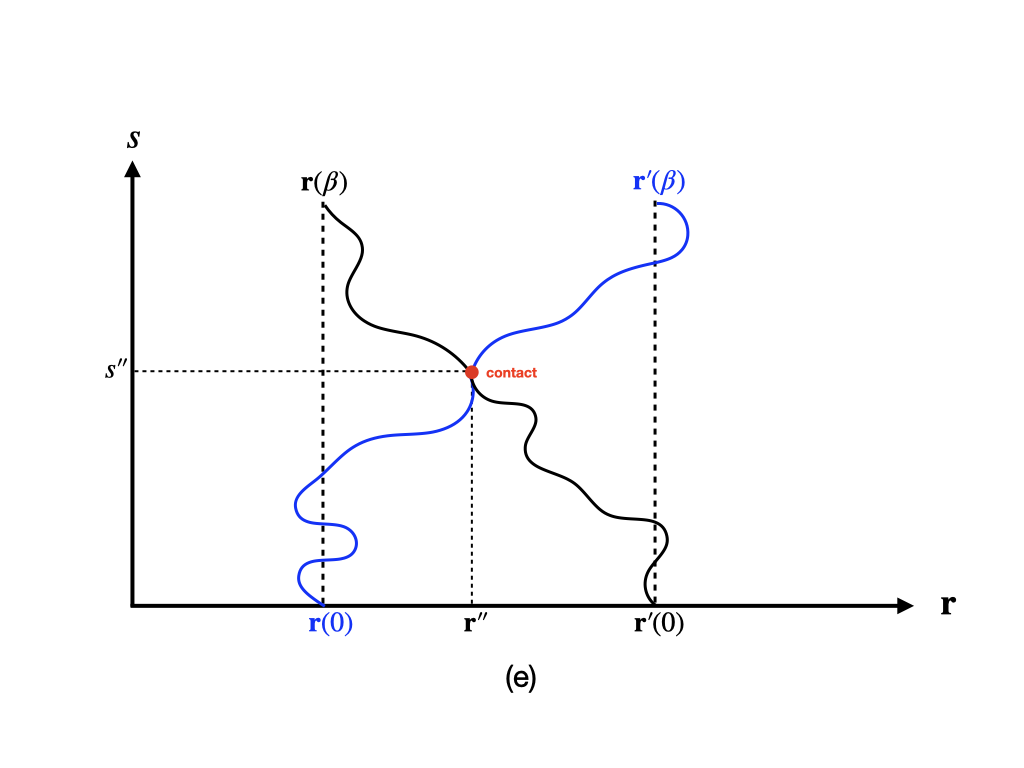} &
\includegraphics[scale=0.15]{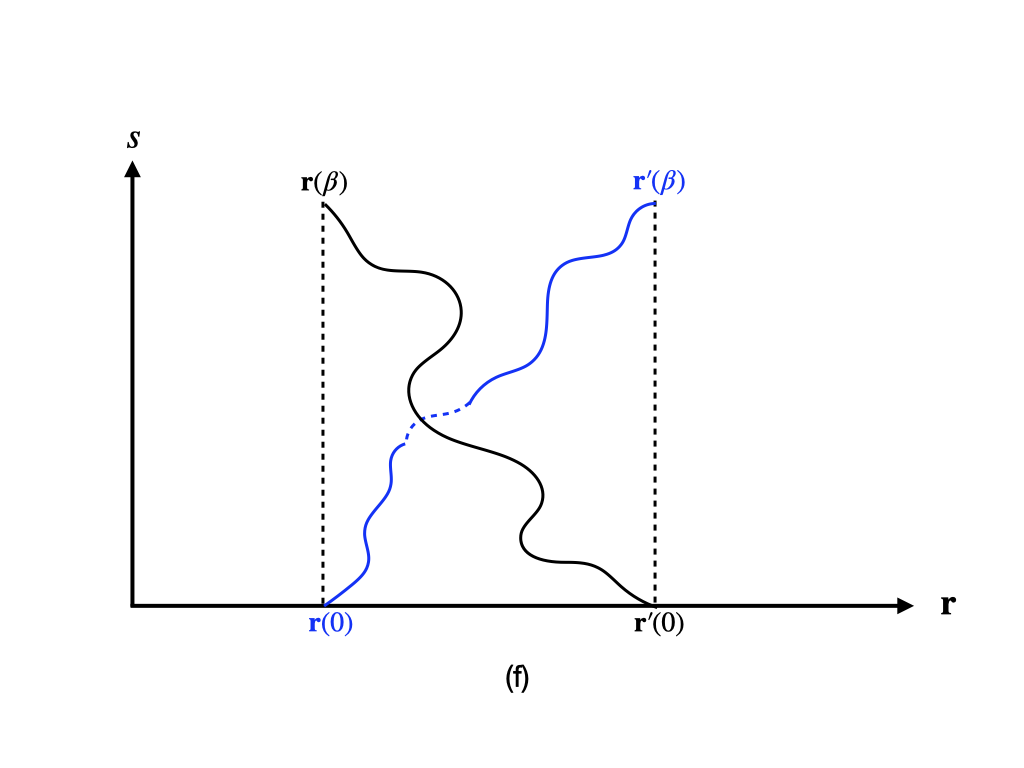} 
\end{tabular}
\caption{Schematics of possible imaginary time trajectories for quantum ``ring polymers''. (a) One possible single particle path. Three dimensional space is collapsed onto the $x$-axis, and the $y$-axis is the inverse thermal energy. (b) One possible configuration of two particles. (c) A configuration of two particles under exchange. (d) Two trajectories that have an overlap at position ${\bf r}^{\prime\prime}$ and imaginary time $s^{\prime\prime}$. (e) Same as panel (d), except the trajectories are cross-paths rather than ring paths. (f) Two crossing trajectories that do not touch; the dashed portion of the blue curve indicates that it passes behind the black curve without contact.} 
\label{fig-contours}
\end{figure}

One must then consider other reasons why the results of references \cite{Thompson2020, LeMaitre2023a, LeMaitre2023b} do not completely agree with HF theory, assuming for the moment that excluded volume is a correct hypothesis. Ignoring fluctuations about the mean field is undoubtably partially responsible, but there is also a topological reason for the discrepancy. Equation (\ref{dens1}) shows that the one-particle density is proportional to the diagonal of the propagator $q({\bf r},{\bf r},\beta)$. This is illustrated in figure \ref{fig-contours}(a), where one possible trajectory that starts at position ${\bf r}$ for $s=0$ returns to the same position ${\bf r}$ for $s=\beta$ -- the probability of finding a particle at position ${\bf r}$ is proportional to $q({\bf r},{\bf r},\beta)$. 
Following reference \cite{Thompson2023}, a two-particle density may be defined which is related to the probability of finding one particle at ${\bf r}$ and another particle at ${\bf r}^\prime$, as illustrated in figure \ref{fig-contours}(b). If the particles are completely uncorrelated, then the two-particle density is just proportional to the product of the propagators
\begin{equation}
n({\bf r},{\bf r}^\prime) \propto  q({\bf r},{\bf r},\beta) q({\bf r}^\prime,{\bf r}^\prime,\beta) .  \label{2dens1}
\end{equation}
Two classical ring polymers which start at ${\bf r}$ and ${\bf r}^\prime$ for $s=0$, respectively, must each return to their starting points at ${\bf r}$ and ${\bf r}^\prime$ for $s=\beta$. Quantum particles however are indistinguishable, and so may exchange final $s=\beta$ locations, as shown in figure \ref{fig-contours}(c). The boson case has been discussed in references \cite{Thompson2023, Ceperley1995, Feynman1953b, Fredrickson2023} for example, where it is shown that boson exchange is equivalent to a classical ring polymer system including microstates where the rings can merge and separate. We show here that it is similarly possible to give a classical description to the fermion case. For fermions, exchange is included in the two-particle density expression (\ref{2dens1}) by changing the sign of the boson case in reference \cite{Thompson2023} to give
\begin{equation}
n({\bf r},{\bf r}^\prime) \propto  q({\bf r},{\bf r},\beta) q({\bf r}^\prime,{\bf r}^\prime,\beta)  - q({\bf r},{\bf r}^\prime,\beta) q({\bf r}^\prime,{\bf r},\beta)     .   \label{2dens2}
\end{equation}
Following the reasoning of Feynman \cite{Feynman1953b}, if the positions ${\bf r}$ and ${\bf r}^\prime$ are very far from each other, there are unlikely to be many trajectories that cross from ${\bf r}$ to ${\bf r}^\prime$ or vice versa. Therefore the probability $q({\bf r},{\bf r}^\prime,\beta)$ will approach zero, and the two-particle expression (\ref{2dens2}) will just be the product of two single particle probabilities, that is, the first term of (\ref{2dens2}). On the other hand, if ${\bf r}$ is sufficiently close to ${\bf r}^\prime$, such that their trajectories overlap, then $q({\bf r},{\bf r}^\prime,\beta)$ will not be small, and equation (\ref{2dens2}) subtracts off these overlaps. In the limit where ${\bf r} = {\bf r}^\prime$, the probability of finding the two particles at the same place ${\bf r} = {\bf r}^\prime$ at the same imaginary time $s = \beta$ is exactly zero from equation (\ref{2dens2}). Thus this equation can be interpreted as enforcing a mean field level excluded volume between the imaginary time trajectories of the two particles. To see this, one can consider the possible imaginary time paths of the particles. Figure \ref{fig-contours}(d) shows a configuration where excluded volume is violated, and the two ring polymer trajectories touch at a point ${\bf r^{\prime\prime}}$ for $s = s^{\prime\prime}$. Figure \ref{fig-contours}(e) shows that this is identical to the conformations of two crossing trajectories, which are subtracted off in the second term of (\ref{2dens2}). In fact all points of trajectory overlap will be subtracted off in equation (\ref{2dens2}), meaning that all excluded volume violating conformations are removed from the possible trajectories. Of course, the second term of (\ref{2dens2}) also subtracts conformations beyond excluded volume ones, as shown in figure \ref{fig-contours}(f). Although such conformations do not violate excluded volume directly, they are topologically inaccessible due to excluded volume. An analogy from real ring polymers is concatenated versus non-concatenated rings. Concatenated rings are joined together like the links of a chain. Although they do not violate excluded volume, such conformations cannot be formed continuously from non-concatenated rings unless excluded volume is broken. SCFT does not capture such topological subtleties, and so will include such forbidden microstates even when excluded volume is included in the model through contact energy penalties of the Edwards-Flory-Huggins type used in references \cite{Thompson2020, LeMaitre2023a, LeMaitre2023b}. In a similar way, configurations such as that shown in figure \ref{fig-contours}(f) cannot be continuously accessed from configurations such as figure \ref{fig-contours}(b) unless that conformation first passes through an excluded volume situation such as figures \ref{fig-contours}(d) and (e). All such topologically inaccessible conformations are removed from the two-particle density $n({\bf r},{\bf r}^\prime)$ by the second term of (\ref{2dens2}).

This argument is valid beyond the two-particle density. For example, the three-particle density, including negative signs only for odd permutations, is
\begin{eqnarray}
n({\bf r},{\bf r}^\prime,{\bf r}^{\prime\prime}) &\propto&  q({\bf r},{\bf r},\beta) q({\bf r}^\prime,{\bf r}^\prime,\beta) q({\bf r}^{\prime\prime},{\bf r}^{\prime\prime},\beta) - q({\bf r},{\bf r},\beta) q({\bf r}^\prime,{\bf r}^{\prime\prime},\beta) q({\bf r}^{\prime\prime},{\bf r}^{\prime},\beta) \nonumber \\
&& - q({\bf r}^\prime,{\bf r}^\prime,\beta) q({\bf r},{\bf r}^{\prime\prime},\beta) q({\bf r}^{\prime\prime},{\bf r},\beta) - q({\bf r}^{\prime\prime},{\bf r}^{\prime\prime},\beta) q({\bf r},{\bf r}^{\prime},\beta) q({\bf r}^{\prime},{\bf r},\beta)  \nonumber \\
&& + q({\bf r},{\bf r}^\prime,\beta) q({\bf r}^\prime,{\bf r}^{\prime\prime},\beta) q({\bf r}^{\prime\prime},{\bf r},\beta) 
+ q({\bf r},{\bf r}^{\prime\prime},\beta) q({\bf r}^{\prime\prime},{\bf r}^{\prime},\beta) q({\bf r}^{\prime},{\bf r},\beta)     \label{3dens1}
\end{eqnarray}
where overlaps between neighbouring pairs of trajectories are subtracted off, and double counting of these overlap subtractions are added back by the last two terms of (\ref{3dens1}). Fermion exchange can be viewed as enforcing both excluded volume and topologically inaccessible conformations. This must be part of the reason why the excluded volume model used in references \cite{Thompson2020, LeMaitre2023a, LeMaitre2023b} only agrees semi-quantitatively with HF -- the Edwards-Flory-Huggins parameter enforces a mean field version of excluded volume, but does not  exclude topologically forbidden conformations. 

Generalizing (\ref{2dens2}) and (\ref{3dens1}) for bosons or fermions, including normalization, gives 
\begin{equation} 
n({\bf r}_1, \ldots, {\bf r}_N,\beta) = \frac{N\sum_{\sigma \in S_N}\prod^N_k q({\bf r}_k, {\bf r}_{\sigma(k)}, \beta)(\pm 1)^\sigma}{\sum_{\sigma \in S_N}\prod^N_k\prod^{N}_{j} \int d{\bf r}_j q({\bf r}_k, {\bf r}_{\sigma(k)}, \beta)(\pm 1)^\sigma}\,  \label{n-particle_exchange_density}
\end{equation}
where $\sum_{\sigma \in S_N}$ is the sum over all possible permutations $\sigma$ up to length $N$; $(1)^{\sigma}$ is for the case of bosons while $(-1)^{\sigma}$ is for fermions. Equation (\ref{n-particle_exchange_density}) is derived by carrying through the anti/symmetrized position basis states $\frac{1}{N_{\pm}!}\prod^N_k\sum_{\sigma \in S_N} (\pm 1)^{\sigma} |{\bf r}_{\sigma(k)}\rangle$, with $N_{\pm}!$ equal to $N!\prod^{N}_{i} n_{i}!$ for bosons ($n$ is the occupation number) and $N!$ for fermions, when deriving the SCFT equations \cite{Thompson2019, LeMaitre2023b}. One can also arrive at equation (\ref{n-particle_exchange_density}) by using the diagrammatic arguments presented for equations (\ref{2dens2}) and (\ref{3dens1}), although it becomes more cumbersome to express for increasing $N$.

We implemented equation (\ref{2dens2}) numerically to show that it gives expected fermion behaviour for atomic systems. Integrating $n({\bf r},{\bf r}^\prime)$ over all ${\bf r}$ and ${\bf r}^\prime$ gives the number of particle pairs in the system, $N(N-1)/2$. We can therefore find the constant of proportionality for (\ref{2dens2}) and write
\begin{equation}
n({\bf r},{\bf r}^\prime,\beta) = \frac{N(N-1)}{2\left[Q(\beta)^2 - Q(2\beta)\right]} \left[q({\bf r},{\bf r},\beta) q({\bf r}^\prime,{\bf r}^\prime,\beta) -q({\bf r},{\bf r}^\prime,\beta) q({\bf r}^\prime,{\bf r},\beta)\right]  . \label{2dens3}
\end{equation}
Still following \cite{Thompson2023}, the one-particle density is obtained by integrating (\ref{2dens3}) over ${\bf r}^\prime$, including a factor $2/(N-1)$ to switch from counting pairs to singlets
\begin{equation}
n({\bf r},\beta) = \frac{N}{\left[Q(\beta)^2 - Q(2\beta)\right]}\left[q({\bf r},{\bf r},\beta) Q(\beta) - q({\bf r},{\bf r},2\beta) \right] . \label{dens3}
\end{equation}
The SCFT equations were solved following the method and details of references \cite{Thompson2019, Thompson2020, LeMaitre2023a}. The diffusion equation (\ref{diff1}) for ring polymers can be more computationally demanding than for linear polymers due to the double spatial dependence on ${\bf r}$ and ${\bf r}_0$. Using a spectral approach however, the numerical cost for solving rings becomes almost equivalent to that for solving linear polymers with a single spatial variable ${\bf r}$ \cite{Qiu2011}. We therefore chose a spectral expansion in terms of Gaussian basis functions which is a common and efficient basis set often used for atomic and molecular systems \cite{Badhan2024}. We self-consistently solved the equations without a Pauli potential, in contrast to references \cite{Thompson2019, LeMaitre2023a, LeMaitre2023b}, using instead equation (\ref{dens3}) in place of (\ref{dens1}) for the beryllium atom. The beryllium atom was chosen for simplicity, since it is the lightest atom with symmetric spin up and spin down electrons that requires the Pauli exclusion principle to give proper shell structure. Since it has only two electrons for each of spin up and spin down, we do not need higher order excluded volume terms, such as the three-body expression (\ref{3dens1}).

\begin{figure}
\centering
\includegraphics[scale=0.5]{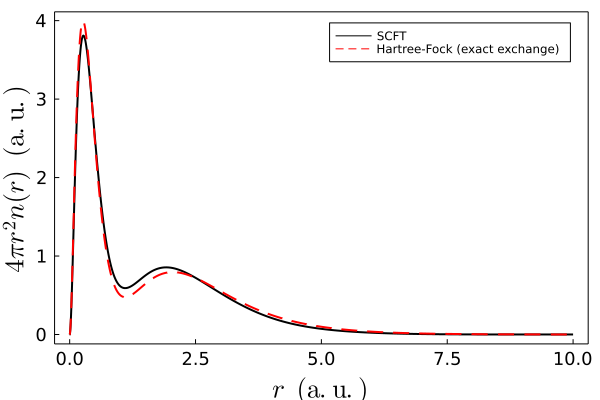} 
\caption{Electron density for the beryllium atom calculated with SCFT (black solid line) and HF (red dashed line). The first principles SCFT result uses no free parameters. Deviation between SCFT and HF is attributed to approximate Coulomb self-interactions in SCFT.} 
\label{fig-exchange}
\end{figure}

Equation (\ref{dens3}) is numerically troublesome because both the numerator and denominator involve differences between extremely large terms that are almost equal -- the terms become identical in the limit $\beta \rightarrow 0$. Feynman and Hibbs noted this sign problem decades ago when they studied the ring polymer partition function \cite{Feynman1965}, and this is part of the reason why bosons are studied with quantum simulation techniques based on ring polymers much more than fermions. We used a brute force solution: replacing double precision floating points in our calculation with very high precision floats. We also chose the smallest $\beta$ value that was still large enough to approach zero temperature. To determine this, we increased $\beta$ until the binding energy and electron density profile stopped changing in any significant way. $\beta=40$ was found to be large enough. We used 75 Gaussian basis functions of the type described in reference \cite{LeMaitre2023a} and converged the fields to a self-consistent tolerance below an L2-norm of $10^{-7}$. We verified that nothing changed by also using over 150 basis functions and a self-consistent tolerance of $10^{-8}$. The electron density is shown in figure \ref{fig-exchange} contrasted with HF theory. 
HF implements exchange exactly but uses a mean field Coulomb potential, and so makes a suitable comparator since we would like to ignore classical correlations for simplicity. We stress that the SCFT result in figure \ref{fig-exchange} is calculated from first principle using only the classical ring polymer partition function with zero free parameters, so despite some differences between the HF and SCFT results in figure \ref{fig-exchange}, the agreement is remarkable. SCFT spontaneously shows shell structure and a total binding energy within less than 6\% of HF without any free parameters, and as already discussed, these features can be explained in entirely classical excluded volume terms. The residual disagreement between SCFT and HF can be accounted for classically as well: despite using a mean field Coulomb electron-electron potential, HF includes electron electrostatic self-interactions exactly whereas SCFT does not. In our calculations, we used a crude Fermi-Amaldi self-interaction correction for the classical electron-electron potential \cite{Ayers2005}. There may be opportunity for improvement here, but this is beyond the scope of this work. 

\section{Summary and Future Outlook} 

Fermion quantum exchange can be incorporated into the quantum-classical isomorphism using thermal trajectory excluded volume. For beryllium, equation (\ref{2dens2}) treats antisymmetric exchange exactly, to the extent that the mean field propagator $q({\bf r},{\bf r},\beta)$ can be determined exactly, by subtracting off forbidden two-body conformations. The microstates that are removed from the one-particle density in (\ref{2dens2}) are those which violate the classical excluded volume of the thermal-space paths directly (two particles occupy the same position at the same imaginary time) or indirectly (the paths are inaccessible topologically). Thus 3D quantum fermion exchange is equivalent to 4D classical excluded volume. This result parallels that of the boson case, which has already been shown to have a quantum-classical mapping between 4D merging and separation of rings and 3D boson exchange \cite{Thompson2023, Ceperley1995, Feynman1953b, Fredrickson2023}. Although an exact agreement between the electron density of beryllium calculated using SCFT via equation (\ref{2dens2}) and HF was not found, the remaining discrepancy can be attributed to self-interactions in the SCFT classical electron-electron potential.

Better agreement between HF and SCFT could be achieved by replacing the Fermi-Amaldi self-interaction correction in the electron-electron potential. Previous work enforcing a Pauli exclusion principle using 4D excluded volume modelled this more approximately as an energy penalty rather than an entropic correction (removal of forbidden paths). The energetic approach required the use of phenomenological parameters analogous to Flory-Huggins parameters in polymer physics, in contrast to the more correct parameter-free entropic approach described here, but the two methods could potentially be combined to improve the self-interaction estimate. The energetic method allows one to keep track of the electron densities of distinguishable electrons, so if one could calibrate the parameters against the entropic method iteratively, one would have a method without free parameters and with a high accuracy self-interaction correction. Once calibrated, such a bootstrapping, or multiscale, approach could be applied to more complex quantum many-body systems since the high precision floating point numerics necessary for equation (\ref{2dens2}) would no longer be required. This combined approach is a possible future direction.

In the mean field SCFT context, keeping track of excluded volume between imaginary time-slices is not necessary. Due to the symmetry of the ring polymer architecture, all interactions between contour values of equal imaginary time are the same. Therefore it is allowable to compute results using only a single contour value, provided the entire polymer contour on either side of the chosen segment contributes following equation (\ref{dens2}). The simplest value is thus $s=\beta$, as chosen in previous work \cite{Thompson2020, LeMaitre2023a, LeMaitre2023b}.

Going beyond the mean field requires comparisons to data which contain classical correlations instead of comparisons to HF which does not. There is much current work on polymer FTS, which includes fluctuations missing in mean field SCFT, and it might be possible to extend the  formalism given here into Langevin simulations, either real or complex, following FTS protocols \cite{Matsen2020, Matsen2021, Fredrickson2023}. In particular, Delaney, Orland and Fredrickson have applied FTS to quantum boson systems \cite{Delaney2020, Fredrickson2022, Fredrickson2023}. However, it is impossible to exclude ring polymer topologically forbidden states using FTS based on SCFT \cite{Matsen2020}, and so a multiscale bootstrap mapping, as described above, might still be needed. On the other hand, Fredrickson et al. apparently avoid this problem by incorporating exchange through a coherent states approach \cite{Delaney2020, Fredrickson2022, Fredrickson2023}. This sacrifices part of the quantum-classical isomorphism, but gives, according to references \cite{Delaney2020, Fredrickson2022, Fredrickson2023}, exact results.

An advantage of keeping a fully classical perspective via the fictitious thermal dimension (imaginary time) using SCFT, is that it can inform quantum foundations. It has been shown that the first principles derivation of the SCFT equations in terms of ring polymers is equivalent to Kohn-Sham density functional theory (DFT) \cite{Thompson2019}. This means that the theorems of DFT give a rigorous one-to-one mapping between an imaginary time ring polymer ontology and predictions of non-relativistic quantum mechanics, including temperature and time-dependent properties \cite{Hohenberg1964, Kohn1965, Mermin1965, Runge1984}. The inclusion of 3D quantum exchange through a 4D classical mapping strengthens this case, especially since 4D excluded volume gives rise spontaneously and without free parameters via equation (\ref{2dens2}) to atomic shell structure, as shown in figure \ref{fig-exchange}, fulfilling the role of the Pauli exclusion principle and fermion exchange.

\begin{acknowledgments}
The authors thank M. W. Matsen for helpful discussions and suggesting the use of direct exchange mechanisms such as equations (\ref{2dens2}) and (\ref{3dens1}). 
\end{acknowledgments}

\bibliography{DFTbibliography}

\end{document}